# Microscopic thickness determination of thin graphite films formed on SiC from quantized oscillation in reflectivity of low-energy electrons

H. Hibino, H. Kageshima, F. Maeda, M. Nagase, Y. Kobayashi, and H. Yamaguchi

*NTT Basic Research Laboratories, NTT Corporation, Atsugi, Kanagawa 243-0198, Japan*

Low-energy electron microscopy (LEEM) was used to measure the reflectivity of low-energy electrons from graphitized SiC(0001). The reflectivity shows distinct quantized oscillations as a function of the electron energy and graphite thickness. Conduction bands in thin graphite films form discrete energy levels whose wave vectors are normal to the surface. Resonance of the incident electrons with these quantized conduction band states enhances electrons to transmit through the film into the SiC substrate, resulting in dips in the reflectivity. The dip positions are well explained using tight-binding and first-principles calculations. The graphite thickness distribution can be determined microscopically from LEEM reflectivity measurements.

Recently, thin graphite films, especially single graphite sheets called graphene, have attracted much attention. This is because they exhibit interesting electronic transport properties, such as field effects and quantum hall effects.[1-3] So far, thin graphite films have been formed in two ways. One is based on processing bulk graphite using oxygen plasma etching,[1,4] but this method cannot provide thin graphite layers with a large area. The other is to anneal SiC surfaces at high temperatures in an ultrahigh vacuum (UHV). Selective sublimation of Si from the substrate results in the graphite films on the surface.[5-10] The graphite films can be processed to fabricate device structures using standard lithographic techniques, and the magnetotransport measurements of the structures have revealed signatures of quantum confinement.[9] This method may provide wide graphite films, which would make it more suitable for device application. However, to use the thin graphite on the SiC substrate for device fabrication, we need a reproducible way of forming graphite films with an intended thickness. For this purpose, it is essential to determine the graphite thickness during various stages of the formation processes. Auger spectroscopy has been used to estimate thickness of graphite formed on SiC.[7] More recently, it has been shown that the number of graphene layers in the graphite film can be determined from the band structure measured using angle-resolved photoemission spectroscopy,[10] but this method also provides only spatially-averaged information. Local thickness distributions are more desirable.

Confinement of electrons in thin films creates quantum well (QW) bound states. QW resonant states can form as well at energies above the confinement potential barrier, because the potential discontinuity scatters electrons quantum-mechanically. To date, photoemission spectroscopy has provided the most direct observation of the QW states, both bound and resonance states, below the Fermi level.[11] Photoemission spectroscopy measurements have revealed that the QW states can cause dramatic quantum size effects on the film properties, such as film stability,[12] magnetic interlayer coupling,[13] and superconductivity.[14] The QW states at discrete energy levels produce peaks in the photoemission energy spectrum. The energy levels of the QW states change with the film thickness. Therefore, the photoemission intensity shows an oscillatory behavior as a function of the electron energy and film thickness. Similarly, reflectivity of low-energy electrons from thin films oscillates depending on the electron beam energy and film thickness.[15-22] The reflectivity oscillation has been commonly understood in terms of the interference between the electron waves reflected from the film surface and the interface between film and substrate.[16,17,19] As a further step in this direction, it has been shown recently that the measured reflectivity oscillations can be explained by Fabry-Pérot type interference of multiply reflected electrons.[21,22] On the other hand, the QW resonant states above the vacuum level should promote the transmission of incident electrons through the thin films, which could cause the quantized oscillation in the electron reflectivity.[15] Although these two interpretations of the reflectivity oscillation using the electron interference and QW states appear to be quite different at first sight, they lead us to the same conclusion, as will be shown in the final part of this paper. In other words, the two interpretations are based on the same physics. Low-energy electron microscopy (LEEM) is currently the best suited method for the reflectivity measurement.[15-22] The film thickness can be determined with atomic layer resolution in the spatial resolution of ~10 nanometers using LEEM.[20]

In this work, we measured the electron reflectivity from thin graphite films formed on SiC(0001) using LEEM. The graphite thickness was determined microscopically from the quantized oscillation in the electron reflectivity. To have a rough image of the reflectivity oscillation, we first obtained the quantization condition for the QW resonant states in a simple square well potential. In the real system, however, the thin films consist of atoms and form the electronic band structures. In thin graphite films, conduction bands whose wave vectors are normal to the film consist of discrete energy levels. These quantized conduction band states are nothing but the QW resonant states. The energy levels depend on the thickness. When the energy of the incident electron coincides with one of the discrete energy levels, the electron resonantly transmits through the film, which reduces the reflectivity. The electron reflectivity oscillates as a function of the electron energy and graphite thickness. To confirm the validity of this scenario, we calculated the quantized conduction band states using the tight-binding approximation. Local minima in the reflectivity agree well with the calculated energy levels. The electron reflectivity measurements using LEEM allow us to observe the graphite thickness distribution in real space, and would greatly contribute to establishing a controllable graphite formation method.

A commercial LEEM instrument (Elmitec LEEM III) was used to investigate the reflectivity from graphitized SiC(0001). We used nominally flat 6H-SiC(0001) wafers (n-type, N-doped, 0.02-0.2 $\Omega$cm). The samples were chemically cleaned, and introduced into UHV through the load lock. The base pressure of the LEEM was less than $5\times10^{-11}$ Torr. The samples were annealed by electron-beam bombardment from the backside. We could not directly measure the sample temperatures using an infrared pyrometer, because the light from the W filament on the sample backside passed through the transparent SiC substrates. Therefore, we measured the sample temperatures using a WRe thermocouple welded to the sample holder. The samples were first outgassed at around 700°C for several hours and then annealed at around 900°C to form the $\sqrt{3}\times\sqrt{3}$ structure. On some samples, we deposited Si to form the 3×3 structure.[6,23] The surface structure changes and graphite formation on 6H-SiC(0001) during annealing in UHV have already been investigated in detail.[5,6] According to Ref. 6, annealing in the range 1000-1050°C transforms the 3×3 structure to the $\sqrt{3}\times\sqrt{3}$ structure. Successive heat treatments above 1080°C induce the gradual development of the $6\sqrt{3}\times6\sqrt{3}$ low-energy electron diffraction (LEED) pattern, and multilayered graphite is formed after annealing at 1400°C.[6] In this work, therefore, the samples were fully graphitized by annealing at high temperatures, typically 1450°C. The pressure during the graphitization was occasionally even higher than $1\times10^{-8}$ Torr. The fully graphitized surfaces seem not to be very sensitive to whether Si deposition is used to prepare the 3×3 structure or not. We also measured the sample surface morphologies *ex-situ* using commercial AFM instruments after the samples had been taken out of vacuum.

Figure 1 shows LEEM images and LEED patterns of SiC(0001) at various stages of the changes in the surface structure from $\sqrt{3}\times\sqrt{3}$ to a surface partially covered with graphite films. These are bright-field (BF) images obtained using the (0,0) beam. The evolution of the LEED patterns is consistent with Ref. 6. The LEED pattern in Fig. 1(b) is typical of the $6\sqrt{3}\times6\sqrt{3}$ structure; bright regions indicated by the dotted ellipses in the LEEM images are $6\sqrt{3}\times6\sqrt{3}$ domains. Figure 1(a) shows a LEEM image of SiC(0001) after annealing at 1060°C. The $6\sqrt{3}\times6\sqrt{3}$ domains nucleated at surface atomic steps on the $\sqrt{3}\times\sqrt{3}$ surface. AFM images of well-prepared $\sqrt{3}\times\sqrt{3}$ surfaces revealed that the step height is always three bilayers. Further annealing at 1230°C caused the surface to be mostly covered with $6\sqrt{3}\times6\sqrt{3}$, as shown in Fig. 1(b). The step shapes changed during the $\sqrt{3}\times\sqrt{3}$-to-$6\sqrt{3}\times6\sqrt{3}$ transformation, indicating the movement of considerable numbers of C and Si atoms. The LEEM image in Fig. 1(c) was obtained after annealing at

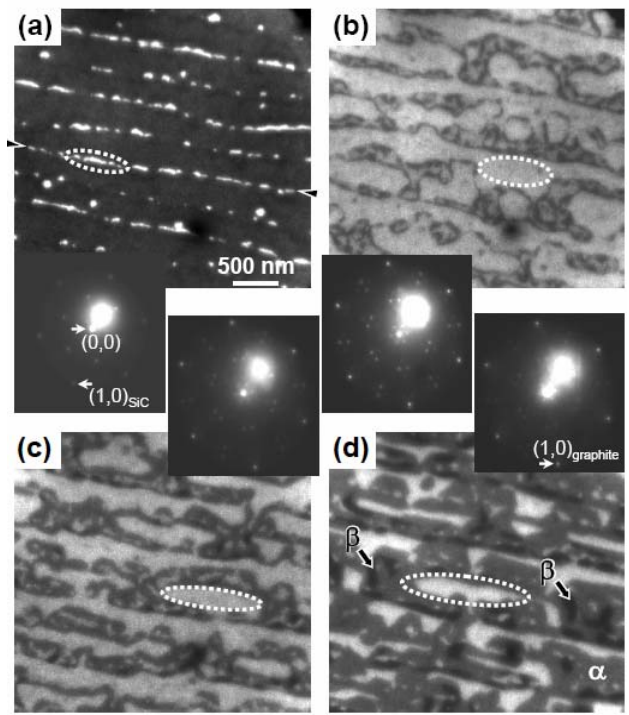

FIG. 1. LEEM images and LEED patterns of 6H-SiC(0001) surfaces after annealing at (a) 1060, (b) 1230, (c) 1320, and (d) 1360°C. These LEEM images were obtained at around 700°C. The electron beam energies are (a) 4.0 and (b)-(e) 3.5 eV. The LEED patterns were obtained at the electron beam energies of (a) 22.0 and (b)-(d) 36.0 eV. In (a), steps are rather straight as indicated by the pair of arrows.

1320°C. It shows the expansion of the dark α phases, which are preferentially formed at the steps. We confirmed using *ex-situ* AFM that the areas in which the α phases gather closely together are holes. Annealing at higher temperatures led to an increase in the intensity of the LEED spots from the graphite, as shown in Fig. 1(d). In the LEEM image of Fig. 1(d), besides the 6√3×6√3 and α phases, β phases with the darkest intensity levels are clearly seen as indicated by the arrows. We also found that the image intensity levels of the different phases change quite differently with electron beam energy. After annealing at higher temperatures, the 6√3×6√3 domains eventually disappeared, and the whole surface was covered with graphite thin films.

Figures 2(a)-2(d) show BF LEEM images of a graphitized SiC(0001) surface at various electron beam energies $E_e$, where $E_e$ is simply the bias difference between the sample and electron gun. These images clearly show that the image intensity levels in different regions change with $E_e$ in different manners. We therefore measure the reflectivity $R$ in areas A-F from the image intensity. The intensities of these areas are plotted in Fig. 2(e). $R$ oscillates as a function of $E_e$ within 0-7 eV. The oscillation periodicity becomes shorter from region A to region F. Therefore, these curves are successively shifted upward in Fig. 2(e). The image intensity level of the β phase in Fig. 1(d) changes with $E_e$ similarly to that of area A. Figure 1 also indicates that A-like β phases appear secondly after the 6√3×6√3 formation during annealing. In Fig. 2(e), therefore, the first and second bottom curves indicate the reflectivity data from the 6√3×6√3 surface and the α phase formed first after the 6√3×6√3 formation, which were measured from different samples.

In Fig. 2(e), the α phase has a dip at around $E_e$ of 2.8 eV. At this energy, a dip and peak appear in turn from this curve to curve F. Furthermore, the number of dips between 0 and 7 eV increases one by one in this order. When we annealed the samples repeatedly at around 1450°C, the regions with shorter periodicities usually expanded. Furthermore, such oscillations were seen only on the graphitized surface. It has been shown that $m$-layer-thick films ($m$=integer) produce ($m$-1) quantum interference peaks, or that is to say $m$ dips, in the reflectivity between successive Bragg peaks.[17,19] As will be shown later, the overall low reflectivity of the graphitized surfaces at $E_e$=0-7 eV corresponds to the conduction band of graphite along the Γ-A direction, normal to the graphite sheet. Graphite has band gaps below 0 eV and above 7 eV. The band gap is the manifestation of the Bragg reflection. On the other hand, the 6√3×6√3 surface has a broad hump in reflectivity between 0 and 12 eV. The reflectivity of the 6√3×6√3 surface is quite different from those of the other phases in Fig. 2(e). These results would mean that the 6√3×6√3 surface has no graphene layer, and that the graphite thickness increases one by one from one layer in the α phase to seven layers in area F. We will verify this by reproducing the positions of the dips in the reflectivity using tight-binding and first-principles calculations.

In this work, we consider that the graphite thickness of the 6√3×6√3 surface is equal to 0. However, there has been a long debate about the atomic structure of 6√3×6√3. Some researchers have reported that this is a reconstructed surface without graphene.[24] But others have reported that the 6√3×6√3 periodicity results from a Moiré fringe between the substrate and graphene layer.[6,25] We cannot exclude the possibility that, because the properties of the graphene in the 6√3×6√3 structure differ considerably from those of the graphite films formed on it, the reflectivity measurements are insensitive to the graphene in the 6√3×6√3 structure. To finalize the debate, further experimental and theoretical studies are essential.

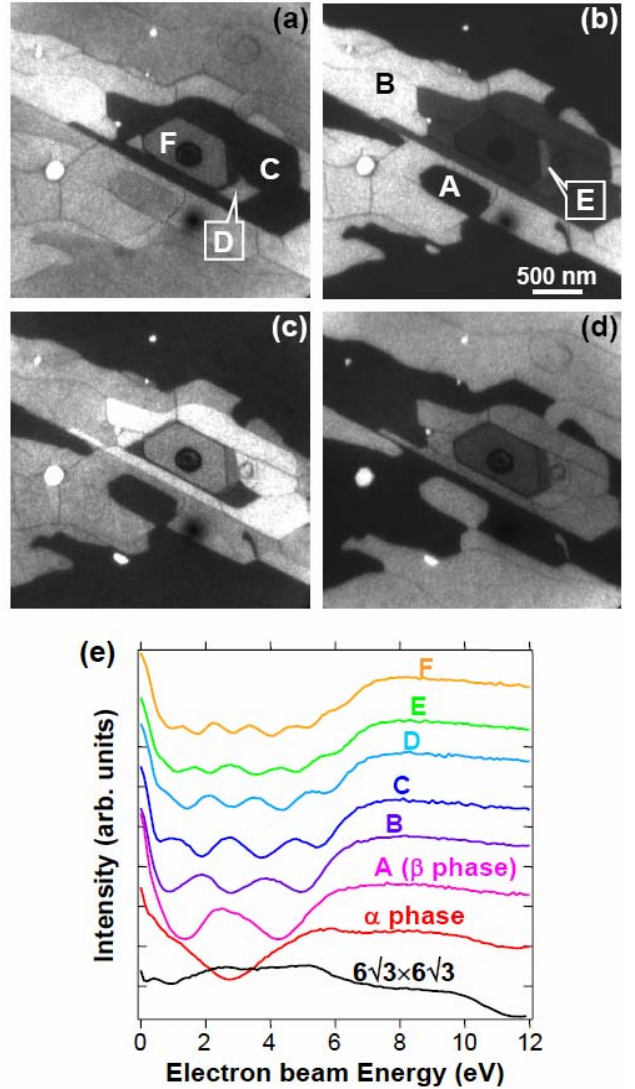

FIG. 2. (a)-(d) LEEM images of a 6H-SiC(0001) surface graphitized at 1460°C. These were obtained after the sample was cooled to room temperature. The electron beam energies are (a) 3.5, (b) 4.0 (c) 4.5, and (d) 5.0 eV, respectively. (e) The reflectivity data from different phases are plotted as a function of the electron beam energy. The data labeled A-E were obtained from area A-E in the LEEM images. The reflectivity data form the 6√3×6√3 and α phases were obtained from the different samples.

The electron energy in materials, $E$, is normally measured from the Fermi level, $E_F$. To convert the electron beam energy $E_e$ to this value, we need to know the work function of the sample surface. The work function is the difference between the vacuum level $E_v$ and $E_F$. The vacuum level position is easily found in the reflectivity data as an energy below which electrons are totally reflected. However, we cannot determine the work function solely using the reflectivity data. Therefore, we measured $E$ from $E_v$. Furthermore, because $E_v - E_F$ could depend on the surface structure, we need a reference vacuum level $E_v^{ref}$ to compare the reflectivity data from different surfaces. We used the reflectivity data with only one peak between 0 and 7 eV as a reference, because the graphite films used in this study have thickness distributions and usually include such regions. Region A in Fig. 2 shows such a reflectivity.

Figure 3(a) shows a reflectivity map as a function of $E - E_v^{ref}$ and graphite thickness. These data were obtained from a couple of different samples. $E_e$=2.8 eV in Fig. 2(e) is equal to $E - E_v^{ref}$ =3.0 eV, at which a dip and peak appear consecutively with increasing thickness. Figure 3(a) indicates that the peak and dip positions systematically change with $E - E_v^{ref}$ and the graphite thickness. This figure looks quite similar to the reported reflectivity oscillations due to the QW resonance of electrons due to the Fabry-Pérot interference effect in MgO thin films on metal substrates.[22]

To interpret the reflectivity data, we first consider a very crude model that illustrates the basic ideas of the oscillation. A free electron with energy $E$ travels over a one-dimensional potential well. The potential is 0 at $x<-d/2$ and $x>d/2$ and $-V$ at $-d/2<x<d/2$. In this case, quantum mechanics teaches us that reflectivity $R$ is

$$R = \frac{V^2(1-\cos 2Kd)}{8K^2k^2 + V^2 - V^2\cos 2Kd}, \quad (1)$$

where the wave vector $k$ at $x<-d/2$ and $x>d/2$ is $k = \sqrt{2mE}/\hbar$ and the wave vector $K$ at $-d/2<x<d/2$ is $K = \sqrt{2m(E+V)}/\hbar$. The reflectivity is equal to 0 at

$$2Kd = 2\pi n, \quad (2)$$

where $n$ is an integer. The potential well is transparent at this quantization condition. This is because the electron forms a QW resonant state in the potential well, and this state promotes electrons to transmit through the potential well.

The reported low-energy electron transmission (LEET) data from bulk graphite show dips at around 7-14 and 21-26 eV.[26] These energy windows correspond to energy gaps in the conduction band of bulk graphite along the Γ-A direction. Our reflectivity data in a wider range of electron beam energy also show broad peaks in these energy windows. This overall correspondence between the transmission and reflectivity data indicates that the energy dependence of the measured reflectivity data should have a strong correlation with the electronic structure of the thin graphite films. Figure 2(e) shows that the electron reflectivity is generally low at $E_e$= 0-7 eV, which is due to the conduction band states at this energy window. The band structure is a reflection of the periodic arrangement of atoms. Therefore, the above crude model using free electrons is insufficient for interpreting the measured reflectivity data quantitatively. Though the conduction band of the bulk graphite is continuous between the Γ and A points, graphite layers with a finite thickness should have discrete energy levels. The dip positions should correspond to them.

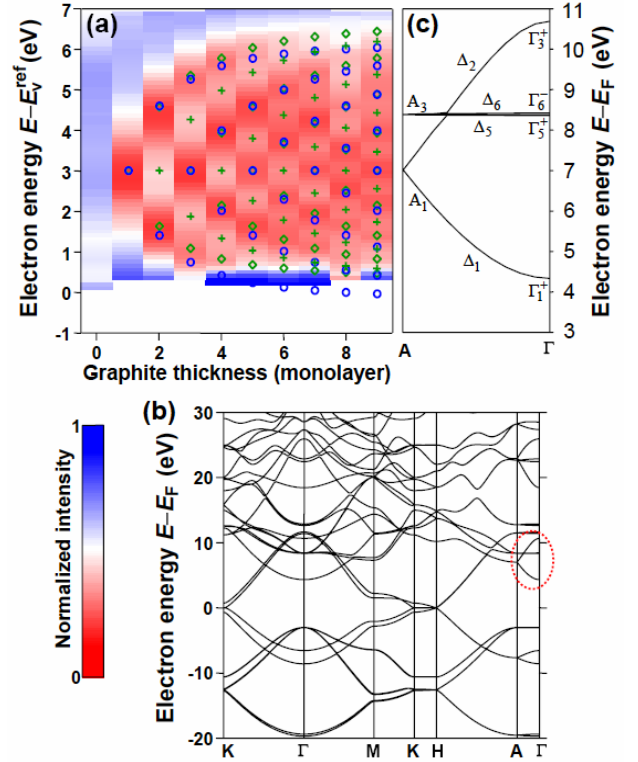

FIG. 3. (a) The reflectivity map as a function of electron energy and graphite thickness. The LEEM observations were done at room temperature. We assume that the graphite thickness is equal to 0 on the 6√3×6√3 surface. In order to compare data from the different samples, the average intensity was obtained at each thickness between $E_e$=0 and 12 eV, and the intensity was normalized by setting the average intensity to 0.5. The intensity is mapped in the red, white, and blue color scheme. The circles indicate the energy level positions determined by the tight-binding calculation, in which the transfer integral of 1.6 eV estimated from the first-principles calculation was used. The diamonds indicate the energy level positions obtained by applying the quantization condition Eq. (3) to the band structure calculated using the first-principles method. The crosses indicate peak positions obtained based on the kinematical scattering theory. (b) The band dispersion relation of the bulk graphite calculated using the first-principles method. (c) The calculated conduction band in the Γ-A direction, which corresponds to the region indicated by the dotted circle in (b).

We evaluate these discrete energy levels using the tight-binding and first-principles calculations. We calculated the band structure of the bulk graphite using a first-principles method based on local density functional theory with ultra-soft pseudopotentials.[27] We used the plane wave basis set up to the cutoff of 25 Ry, 4×4×2 $k$-points, Vanderbilt-type ultrasoft pseudopotentials,[28] and the generalized gradient approximation exchange-correlation functional proposed by Perdew et al.[29] However, we did not succeed in determining the conduction bands of thin graphite films using the first-principles method, because the periodic boundary condition causes free electrons in the vacuum to form artificial resonant multi-bands and these bands overlap with the conduction bands of the thin films. Therefore, we estimated the conduction band levels using the tight-binding calculation, in which the molecular orbitals on the graphite sheets are used as the base set. Fretigny et al. successfully calculated conduction bands in graphite using the linear combination of atomic orbital base first-principles method, which ensures that the tight-binding scheme is a good approximation for discussing the conduction band energy dispersion relations perpendicular to the graphite layers.[30] In the simplest tight-binding scheme, the bulk band dispersion is described as $E = \varepsilon - 2t\cos(Ka)$, where $\varepsilon$ is the energy of the band center, $t$ is the transfer integral, and $a$ is the interlayer distance. The band width is $4t$.

We estimated the band width from the first-principles calculation results. Figures 3(b) shows the calculated band structure of bulk graphite. We clearly see two conduction bands that span several electron volts along the Γ-A direction. These bands well correspond to the high-intensity energy windows in the reported LEET spectra,[26] considering the work function of bulk graphite 4.6 eV.[31] The conduction band of concern is indicated by the dotted circle in Fig. 3(b) and is enlarged in Fig. 3(c). The band width is estimated to be 6.4 V. The reported band calculation gives a similar value.[32] Energy levels of $m$-layer-thick films are eigenvalues $E = \varepsilon - 2t\cos[\pi n/(m+1)]$ of the $m \times m$ tight-binding secular determinant, where $n$=1 to $m$.[33] Circles in Fig. 3 are the energy levels calculated using $\varepsilon = E_v^{ref} + 3.0$ eV and $4t$=6.4 eV. The calculation reproduces the dip positions in the reflectivity fairly well. The number of graphene layers in the graphite film is counted using the number of dips. The reflectivity map in Fig. 3(a) is almost symmetric with respect to the line of $E - E_v^{ref}$ =3.0 eV. At this energy, the reflectivity oscillates with the thickness periodicity of two monolayers. Two monolayers is the unit cell length of the bulk graphite along the $c$ axis, normal to the graphite sheet, $c_0$. From the QW resonance point of view, the wave vector at this energy is $\pi/c_0$, which lies at the A point (Brillouin zone boundary) of the bulk graphite. This is also consistent with the idea that the reflectivity oscillation is related to the graphite band structure.

The first-principles calculation indicates that the conduction band in the Γ-A direction starts from the Γ point at $E_F$+4.3 eV, reaches the A point at $E_F$+7.0 eV, and returns to the Γ point at $E_F$+10.7 eV. The widths of the unfolded and folded bands are different. This difference is visible in Fig. 3. The dip positions predicted by the tight-binding calculation are perfectly symmetric with respect to the line of $E - E_v^{ref}$ =3.0 eV. However, the energy positions in the experimental data are a little larger than the predictions at $E - E_v^{ref}$ <3.0 eV, indicating that the lower unfolded band width is smaller than the higher folded band width. In the simplest tight-binding scheme, the bulk band dispersion is written as $E = \varepsilon - 2t\cos(Ka)$. Substitution of $K = \pi n/(m+1)a$ into the bulk band dispersion provides the discrete energy levels, at which dips appear in the reflectivity. This wave vector satisfies the quantization condition

$$2K(m+1)a = 2\pi n. \quad (3)$$

A comparison between Eqs. (2) and (3) indicates that, regarding the quantization condition only, $m$-layer-thick films act as a square potential well with the thickness of $(m+1)a$ and that Eqs. (2) and (3) are basically the same. Applying Eq. (3) to the calculated band structure in Fig. 3(c) should allow us to reproduce the dip positions more precisely. The diamonds in Fig. 3(a), which are energy levels calculated in this way, confirm this.

To finalize our scenario, in which the quantized oscillation is due to the resonance of the incident electrons with the quantized conduction band states, we have to check the symmetry of the band, because it determines whether the band can couple to the electron waves outside the film or not. Experimentally, the reported LEET data indicate that the LEET intensity is high at 0-6 eV, which means that the conduction band shown in Fig. 3(c) can be accessed from the vacuum side. In fact, this band is known as the interlayer band and has attracted much attention, especially because it plays an important role in understanding the electronic structure of the graphite intercalation compounds.[34-36] The previous theoretical works on the band structure of graphite have shown that this conduction band has the symmetries labeled in Fig. 3(c).[37,38] It is known that elastic electron scattering is sensitive to bands with the $\Delta_1$ symmetry.[15] Matrix elements for the irreducible representations along $\Delta$ indicate that $\Delta_1$ and $\Delta_2$ are the same for the primitive translations but different in sign for the nonprimitive translations which include a translation by a half unit along the $c$ axis.[39] Therefore, we naturally think that elastic electron scattering is sensitive to both $\Delta_1$ and $\Delta_2$ bands. Furthermore, reported first-principles calculation results have shown that the charge-density contributions of the wave functions at $\Gamma_1^+$ and $\Gamma_3^+$ involve relatively flat features localized between atomic planes,[31] which should strongly couple with the electron waves outside the film. Both the experimental and theoretical considerations indicate that the quantized states originated from the conduction band in Fig. 3(b) are observable by the reflectivity.

On the other hand, we do not fully understand the role the substrate plays in the quantized oscillation yet. Bulk 6H-SiC has a complicated band structure along the $\Gamma$-A direction.[40] Furthermore, the measured reflectivity data from the $\sqrt{3}\times\sqrt{3}$ and $6\sqrt{3}\times6\sqrt{3}$ surfaces are quite different. The reflectivity data from SiC sensitively depend on the surface structure. We are unable to interpret these reflectivity data based on the bulk band structure as straightforwardly as in the case of graphite. Therefore, we note only one role of the substrate here. As seen in Fig. 2(e), the reflectivity from the $6\sqrt{3}\times6\sqrt{3}$ surface is higher than that from the surfaces covered with graphite at around 0-6 eV, which would be essential for observing the quantized oscillation.

The calculated conduction band level at the A point is $\varepsilon=E_F+7.0$ eV, which means $E_v^{ref} - E_F$ =4.0 eV. The reported vacuum level of 6H-SiC is 3.8 eV above the conduction band minimum $E_c$.[41] N-doped, $n$-type 6H-SiC with the resistivity of 0.02-0.2 $\Omega$cm has the Fermi level in the band gap, just below $E_c$.[8] Furthermore, the measured work function of the graphite is 4.6 eV.[32] Therefore, the 2-layer-thick graphite film should have a work function between 3.8 and 4.6 eV. This is consistent with $E_v^{ref} - E_F$ =4.0 eV.

So far, we have demonstrated that the resonance of the incident electrons with the conduction band states in graphite films leads to dips in the electron reflectivity. On the other hand, the reflectivity oscillation has been commonly understood in terms of the interference between the electron waves reflected from the film surface and from the interface between film and substrate.[16,17,19] Here, we show that these two interpretations lead us to the same conclusion using a very simple model. In the kinematical scattering theory, the scattering intensity from a crystal with a finite size is proportional to the Laue function. In order to reproduce the measured quantized oscillation in Fig. 3(a), we obtained peak and dip positions using the Laue function of $m$-layer-thick films. Because we measured the (0,0) beam intensity, the problem is reduced to one dimension. We simply assume that a monolayer-thick film on the substrate has two scattering centers separated by distance $a$, which may be regarded as the surface and interface, and that a $m$-layer-thick film has $m+1$ scattering centers due to the periodic potential in the film. Therefore, the Laue function $L(K)$ of the $m$-layer-thick film is written as $L(K) = \sin^2[(m+1)Ka/2]/\sin^2(Ka/2)$. $L(K)$ has main peaks at $K = 2\pi n/a$, which constitute the reciprocal lattice, and ($m$-1) sub-peaks ($m$ dips) between two adjacent reciprocal lattice points. In the backscattering geometry, electron beams with $K = \pi n/a$ produce Bragg peaks, and ($m$-1) sub-peaks appear between successive Bragg peaks. This is consistent with the previously mentioned relationship between the number of layers and number of quantized peaks.[17,19] The backscattered intensity has dips at $2K(m+1)a = 2\pi n$, which is the same as Eq. (3). We also obtained $K$ values which produces peaks in the backscattered intensity. Cross symbols in Fig. 3(a) are energy levels calculated from these $K$ values using the band dispersion in Fig. 3(c). The calculations well reproduce the measured peak positions. Two interpretations – resonant transmission through the QW states produces dips in the reflectivity and the interference of the electron waves reflected from the surface and interface produces peaks in the reflectivity – look quite different, but are essentially the same in meaning.

In summary, we measured the reflectivity of low-energy electrons from the thin graphite layers using LEEM. The reflectivity oscillates as a function of the electron beam energy and graphite thickness. The reflectivity decreases when the electron beam resonantly transmits through the quantized conduction band states of thin graphite films into the SiC substrate. The reflectivity oscillation enables us to determine the graphite thickness with atomic layer resolution. *In-situ* microscopic determination of the graphite thickness using LEEM would greatly contribute to establishing a controllable way of forming wide graphite films with an intended thickness.